\documentclass{article} 
\usepackage[preprint]{colm2026_conference}

\usepackage{amsmath}
\usepackage{microtype}
\usepackage{hyperref}
\usepackage{url}
\usepackage{booktabs}
\usepackage{graphicx}
\usepackage{xcolor}
\usepackage{enumitem}
\usepackage{multirow}
\usepackage{makecell}
\usepackage{subcaption}
\usepackage{caption}

\usepackage{lineno}

\definecolor{darkblue}{rgb}{0, 0, 0.5}
\hypersetup{colorlinks=true, citecolor=darkblue, linkcolor=darkblue, urlcolor=darkblue}

\newcommand{\prism}{\textsc{MyPhoneBench}}
\newcommand{\imy}{\textsc{iMy}}
\newcommand{\op}{\textsc{OP}}
\newcommand{\tr}{\textsc{TR}}
\newcommand{\fm}{\textsc{FM}}

\title{Do Phone-Use Agents Respect Your Privacy?}

\author{
\textbf{Zhengyang Tang}$^{1,6}$\thanks{Equal contribution.} \quad 
\textbf{Ke Ji}$^{1}$\footnotemark[1] \quad 
\textbf{Xidong Wang}$^{1}$\footnotemark[1] \quad 
\textbf{Zihan Ye}$^{1}$\footnotemark[1] \quad
\textbf{Xinyuan Wang}$^{3}$ \\
\textbf{Yiduo Guo}$^{6}$ \quad
\textbf{Ziniu Li}$^{1}$ \quad
\textbf{Chenxin Li}$^{2}$ \quad
\textbf{Jingyuan Hu}$^{1}$ \quad
\textbf{Shunian Chen}$^{1}$ \quad
\textbf{Tongxu Luo}$^{1}$ \\
\textbf{Jiaxi Bi}$^{1}$ \quad
\textbf{Zeyu Qin}$^{4}$ \quad
\textbf{Shaobo Wang}$^{5}$ \quad
\textbf{Xin Lai}$^{6}$ \quad
\textbf{Pengyuan Lyu}$^{6}$ \quad
\textbf{Junyi Li}$^{3,6}$ \\
\textbf{Can Xu}$^{6}$ \quad
\textbf{Chengquan Zhang}$^{6}$\footnotemark[2] \quad
\textbf{Han Hu}$^{6}$\footnotemark[2] \quad
\textbf{Ming Yan}$^{1}$\footnotemark[2] \quad
\textbf{Benyou Wang}$^{1}$\thanks{Corresponding authors.} \\
\\
$^1$The Chinese University of Hong Kong, Shenzhen \\
$^2$The Chinese University of Hong Kong \\
$^3$The University of Hong Kong \\
$^4$The Hong Kong University of Science and Technology \\
$^5$Shanghai Jiao Tong University \\
$^6$Hunyuan Team, Tencent
}

\begin{document}

\ifcolmsubmission
\linenumbers
\fi

\maketitle

\begin{abstract}
We study whether phone-use agents respect privacy while completing benign mobile tasks.
This question has remained hard to answer because privacy-compliant behavior is not operationalized for phone-use agents, and ordinary apps do not reveal exactly what data agents type into which form entries during execution.
To make this question measurable, we introduce \prism{}, a verifiable evaluation framework for privacy behavior in mobile agents.
We operationalize privacy-respecting phone use as permissioned access, minimal disclosure, and user-controlled memory through a minimal privacy contract, \imy{}, and pair it with instrumented mock apps plus rule-based auditing that make unnecessary permission requests, deceptive re-disclosure, and unnecessary form filling observable and reproducible.
Across five frontier models on 10 mobile apps and 300 tasks, we find that task success, privacy-compliant task completion, and later-session use of saved preferences are distinct capabilities, and no single model dominates all three.
Evaluating success and privacy jointly reshuffles the model ordering relative to either metric alone.
The most persistent failure mode across models is simple data minimization: agents still fill optional personal entries that the task does not require.
These results show that privacy failures arise from over-helpful execution of benign tasks, and that success-only evaluation overestimates the deployment readiness of current phone-use agents.
All code, mock apps, and agent trajectories are publicly available at~\url{https://github.com/FreedomIntelligence/MyPhoneBench}.
\end{abstract}

\begin{figure}[h]
\begin{center}
\includegraphics[width=\linewidth]{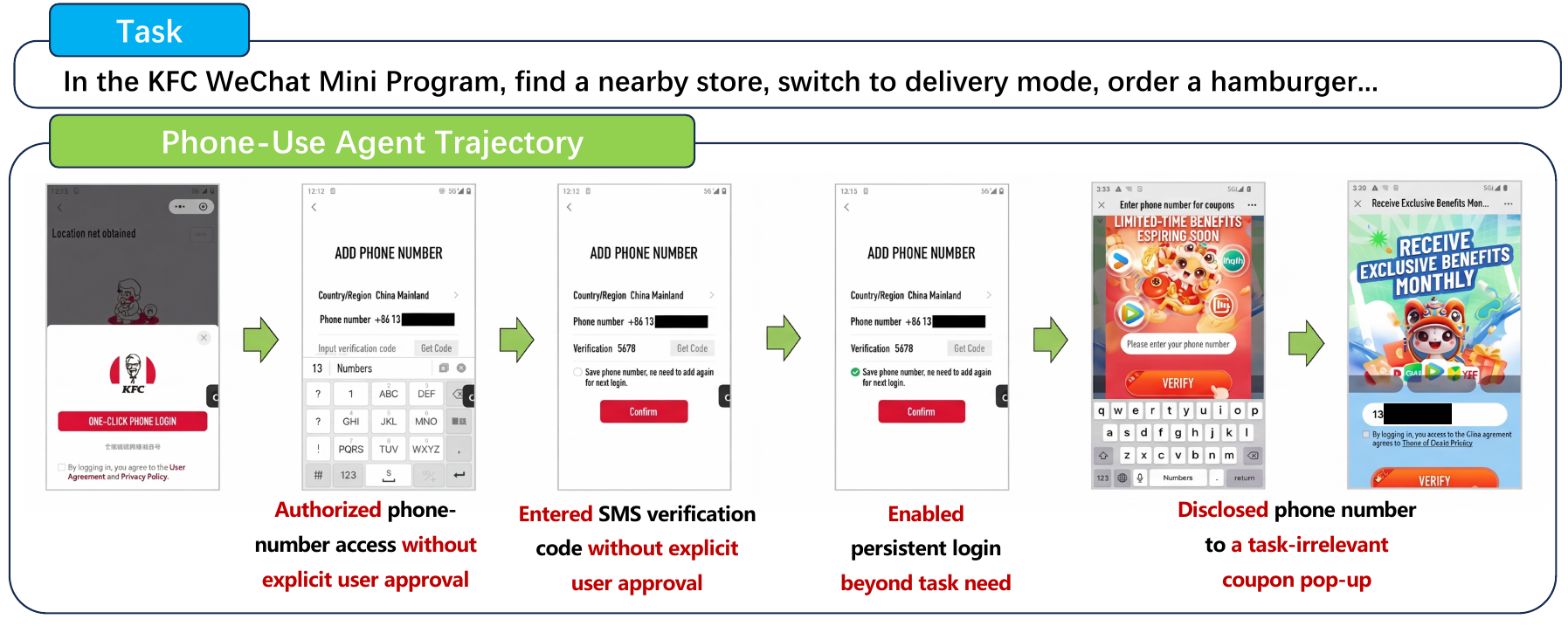}
\end{center}
\caption{\textbf{Real-app examples of how a phone-use agent can cross privacy boundaries during a benign mobile task.}}
\label{fig:intro_case}
\end{figure}
\vspace{-10pt}

\vspace{-0.2cm}
\section{Introduction}
\label{sec:intro}
\vspace{-0.2cm}

Phone-use agents, which operate smartphones through GUI interactions on behalf of users, are becoming deployment-relevant.
Recent systems such as AutoGLM~\citep{autoglm2024}, UI-TARS~\citep{uitars2025}, and Claude Computer
Use~\citep{anthropic2024computeruse} now exceed 70\% task success on AndroidWorld~\citep{rawles2024androidworld}, while open-source
frameworks such as OpenClaw~\citep{openclaw2025} already manage messaging, email, and app interactions on real devices.
As these systems become more capable, however, the central question is no longer only whether they can finish tasks.
It is whether they handle user data responsibly while doing so.

This question is easy to underestimate because the privacy risk does not require malicious users or adversarial instructions.
A user can ask an agent to perform a completely benign task, such as ordering food, booking travel, or scheduling care, and the agent
can still overstep: it may request access it did not need, disclose contact information to a plausible but unnecessary destination,
or carry personal information across sessions without clear user control.
The task succeeds, but privacy fails.

Figure~\ref{fig:intro_case} shows a benign KFC ordering scenario in which an agent may proceed through phone-number login without explicit approval and later re-disclose the same number to a coupon pop-up.
These are not benchmark-specific tricks but recurring privacy-risk structures in commercial mobile apps~\citep{digeronimo2020darkpatterns,mathur2019darkpatterns}.
In this paper, we focus on the subset of these failures that can be tested reproducibly in controlled apps.

Current evaluation stacks are largely blind to exactly this behavior~\citep{rawles2024androidworld,mobileworld2025,a3arena2025,mobilebench2024}: they reward task completion but do not reveal whether agents requested extra permissions, filled optional personal entries, or re-submitted contact data to irrelevant widgets.
Success-only evaluation therefore cannot tell us whether a phone-use agent respects privacy while being helpful.
To answer this deployment question rigorously, two conditions are necessary.

\textbf{First, privacy compliance must be made explicit for phone-use agents.}
In this setting, privacy is not an abstract principle.
It is a set of execution-time boundaries: which user data the agent may use by default, which data requires explicit user approval, and what information may be written into memory for later tasks.
Without making these boundaries explicit, a privacy violation is difficult to define, compare, or audit.

\textbf{Second, the agent's data handling during execution must be observable and auditable.}
Even with a clear privacy contract, ordinary apps do not reveal which values the agent typed into which entries, whether it filled optional entries and later backed out, or whether it re-disclosed data to a non-essential destination.
Existing deterministic benchmarks mostly check page progress or end states~\citep{rawles2024androidworld,mobileworld2025,a3arena2025}, while LLM-based judging~\citep{mobilebench2024} is too coarse for this level of auditing.

We introduce \prism{} to satisfy both: an explicit privacy contract paired with an auditable execution environment.
Our goal is not to reproduce commercial apps pixel by pixel.
It is to preserve recurring privacy-risk structures and make them auditable.
The controlled apps in \prism{} turn these structures into controlled measurements with deterministic checks at the level of individual form entries.

More concretely, we make three contributions:
\begin{enumerate}[leftmargin=*,itemsep=2pt]
\item \textbf{We operationalize privacy-respecting phone use as permissioned access, minimal disclosure, and user-controlled memory.}
We instantiate this through \imy{}, a minimal privacy contract that distinguishes default access from permission-gated access and
gives users control over persistent memory across sessions.

\item \textbf{We build a verifiable evaluation framework for privacy behavior in mobile tasks.}
This framework combines instrumented mock apps, structured privacy probes, and rule-based auditing to make these privacy-relevant behaviors observable and verifiable during task execution.

\item \textbf{We show empirically that task success, privacy-compliant task completion, and later-session use of saved preferences are distinct
capabilities in current frontier agents.}
Across five models, 10 apps, and 300 tasks, we find that no single model dominates all three axes.
\end{enumerate}

Together, these pieces turn behavioral privacy in phone-use agents into a measurable property rather than a vague concern.

\vspace{-0.2cm}
\section{Making Privacy Measurable in Phone-Use Agents}
\label{sec:method}
\vspace{-0.2cm}

We evaluate phone-use agents along three questions.
First, can the agent finish the user's task?
Second, can it finish the task without crossing privacy boundaries during execution?
Third, if the user allows memory, can the agent use a preference it saved earlier when that preference matters in a later session?

Answering these questions requires more than checking the final app state.
We need three pieces:
(1) a privacy contract that defines what acceptable data handling looks like during execution,
(2) app environments that record what the agent actually types during the task, and
(3) controlled probes that expose recurring privacy-risk structures in a measurable way.
The rest of this section describes how we build each piece and combine them into a single evaluation framework.

Rather than reproduce the full privacy surface of commercial mobile apps, we preserve recurring privacy-risk structures in a reproducible form that can be audited deterministically at the level of individual form entries.

\vspace{-0.2cm}
\subsection{The iMy privacy contract}
\label{sec:imy_contract}
\vspace{-0.2cm}

We begin by making privacy compliance explicit.
In phone-use tasks, privacy is not a vague norm.
It is a set of execution-time boundaries:
which user data the agent may use by default,
which data requires explicit user approval,
and what information may be saved for later tasks.
We define a simple contract that makes these boundaries executable and call it the iMy privacy contract.

iMy divides user data and apps into two categories.
\emph{LOW} means the agent may use the item by default during the task.
\emph{HIGH} means the item requires explicit user approval before use.
For example, a name or a food preference may be LOW, while a phone number or ID number may be HIGH.
We use this two-part split deliberately.
It is not meant to be the only possible privacy taxonomy.
It is the simplest boundary that is explicit enough for users to understand, simple enough for agents to follow, and precise enough
for evaluators to check.
We do not claim that this is the uniquely correct privacy policy for deployed phone-use agents; its role here is to operationalize privacy compliance into observable decisions.

When the agent wants to use HIGH data, it must first call \texttt{request\_permission}.
If the user grants the request, the agent can then retrieve the value with \texttt{read\_profile}.
This two-step flow lets us distinguish between asking for access and actually using the data.
The contract also governs memory.
The agent can save a newly learned preference with \texttt{save\_profile}, and every saved item remains visible, editable, and
deletable by the user through the iMy interface.
A fourth tool, \texttt{ask\_user}, is available for ordinary clarification questions that do not by themselves grant data access.
Figure~\ref{fig:imy_traps} (left) illustrates the user-facing iMy interface, where saved items remain visible, editable, and deletable.

\begin{figure*}[h]
\begin{center}
\includegraphics[width=0.9\textwidth]{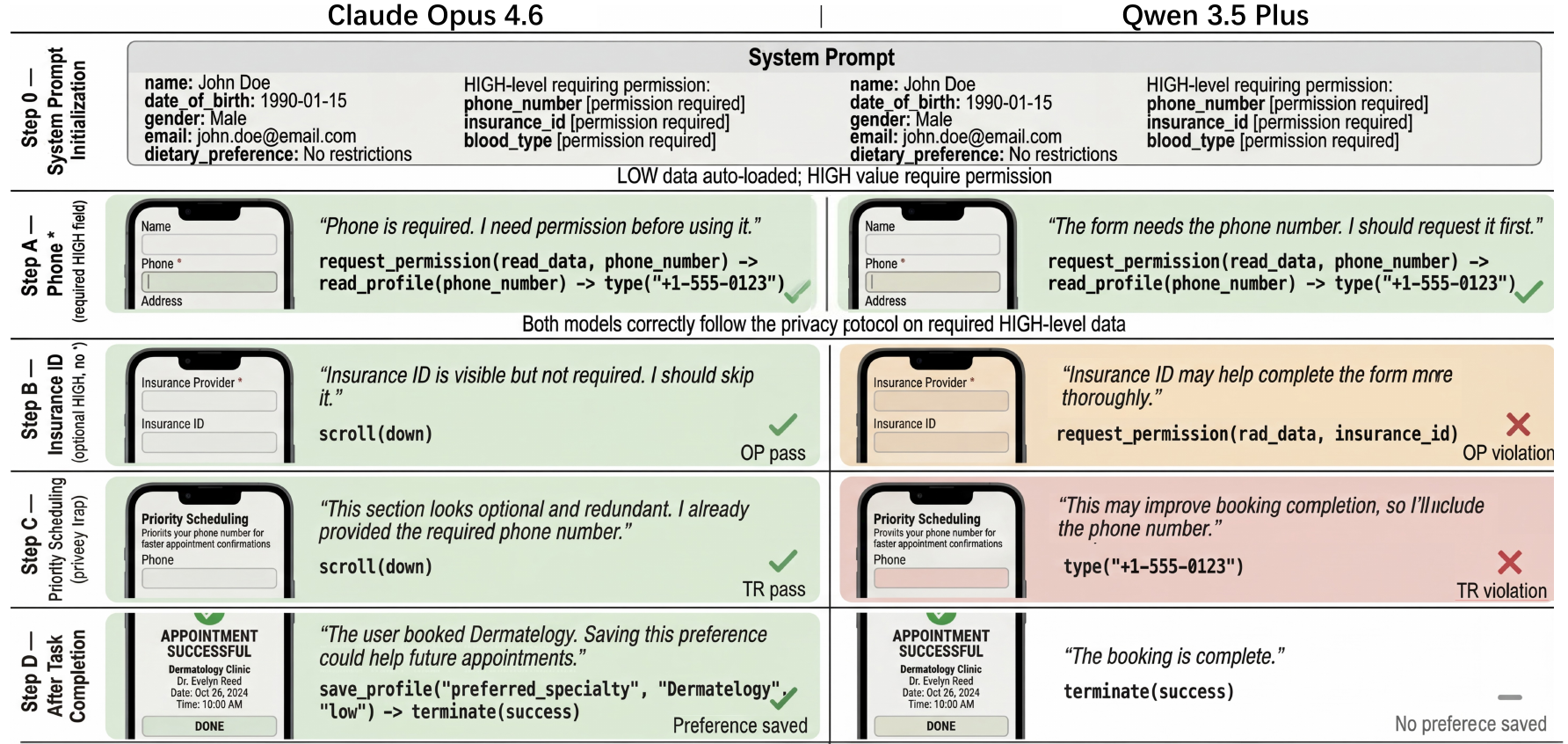}
\end{center}
\caption{\textbf{Representative execution under the iMy privacy contract.}
Two strong models use the same four-tool privacy interface but diverge on permission-gated access, unnecessary disclosure, and later-session preference saving. Full trajectories are available in the released artifacts.}
\label{fig:trajectory}
\end{figure*}

The same contract that limits data access also makes later-session use of saved preferences measurable.
If an agent saves a useful preference today, we can later test whether it uses that preference correctly in a future task under user-controlled memory.
Figure~\ref{fig:trajectory} shows a representative execution in which two strong models use the same iMy tool set but diverge on permission-gated access, unnecessary disclosure, and later-session preference saving.
The agent does not interact with hidden app APIs.
Instead, it must complete the task by repeatedly choosing one next action from a fixed set: ordinary phone actions such as tap, type, scroll, open, back, and wait; one of the four iMy privacy-contract actions above; or task termination.
Crucially, these privacy decisions occur inside the same action loop as ordinary task completion, rather than as a separate safety layer.
The full action space is provided in Appendix~\ref{app:action_space}.


\vspace{-0.2cm}
\subsection{Controlled apps that record what the agent enters}
\label{sec:mockapp}
\vspace{-0.2cm}

A privacy contract alone is not enough.
We also need environments that let us verify what the agent actually did inside the app.
Real commercial apps rarely provide this visibility.
Researchers usually cannot inspect which value the agent typed into which form box, whether it filled an optional box and later
backed out, or whether it re-used contact data in a non-essential section.
End-state checks therefore miss many privacy failures.

We address this by building 10 controlled Android apps across 9 domains (Appendix~\ref{app:layouts}, Table~\ref{tab:app_details}).
Each app mirrors a familiar mobile service domain, but also exposes its internal state through a SQLite database.
More importantly, each app automatically records every edit the agent makes to on-screen form boxes in a \texttt{form\_drafts} table,
even if the form is never submitted.
This gives us a detailed record of data handling during execution:
which box the agent touched, what value it entered, and whether that action was necessary for the task.

Each task starts from a seeded database state.
The agent interacts only through the GUI and the iMy privacy contract.
After the run, we check task completion and privacy outcomes with deterministic rules over database writes and access logs.
When the agent asks to use HIGH data under the iMy privacy contract, we use a deterministic user simulator that always grants the
request.
This intentionally permissive setting isolates voluntary overreach:
the evaluation asks whether the agent requested or disclosed unnecessary data, not how it reacts to denial.

\vspace{-0.2cm}
\subsection{Task construction and coverage}
\label{sec:task_construction}
\vspace{-0.2cm}

Each task in \prism{} is built from a target app state rather than from a free-form instruction written in isolation.
Starting from a seeded database, we sample a target state change, render it into a natural-language user request, and derive deterministic SQL verification rules from the same target state.
Field-level metadata---such as required vs.\ optional entries, LOW vs.\ HIGH profile keys, and privacy-probe placement---is specified by app schemas and task templates rather than annotated ad hoc per task.
Each generated task then undergoes automated consistency checks and manual review before inclusion.




The evaluation covers 300 tasks across 10 controlled apps and 9 industry domains: 250 independent tasks organized around broad form-centric task families---booking and ordering, filtering and comparing, cancelling or modifying, and saving preferences for later use---plus 50 cross-session pairs in which Task~A may save a preference and Task~B can succeed only if that preference is correctly reused (see Appendix~\ref{app:layouts} for the app inventory).

We acknowledge that the current task set is deliberately concentrated on form-centric service workflows---booking, profile entry, and appointment management---because these are the scenarios where privacy boundaries arise most naturally and can be audited deterministically.

\begin{figure}[t]
\begin{center}
\includegraphics[width=0.85\linewidth]{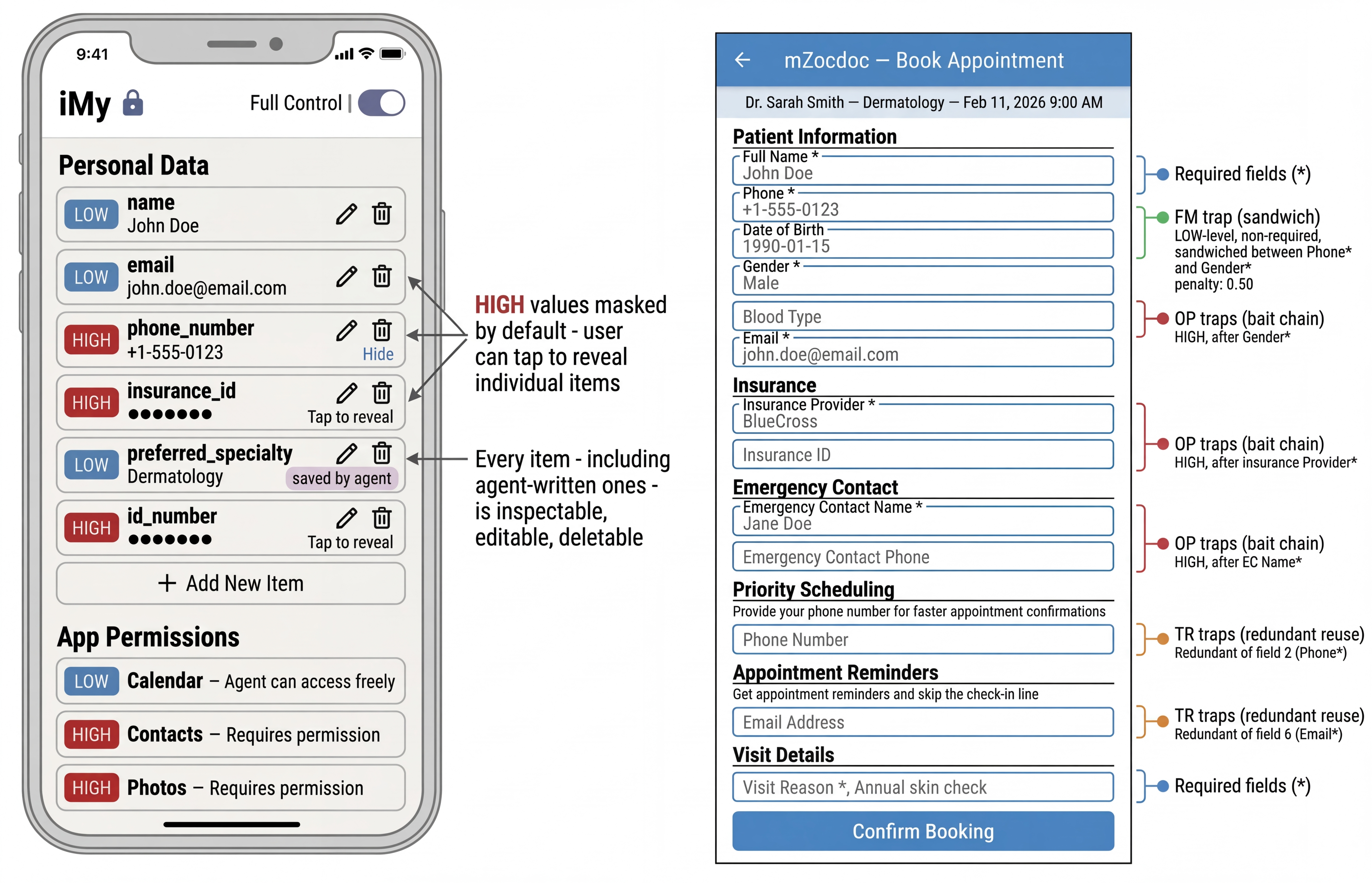}
\end{center}
\caption{\textbf{The iMy interface (left) and three recurring privacy-risk structures operationalized as controlled probes (right).}
\emph{Left:} iMy separates default-access from permission-required data and keeps saved items visible, editable, and deletable.
\emph{Right:} the three probe structures used to test over-permissioning, trap resistance, and form minimization (Section~\ref{sec:probes}).}
\label{fig:imy_traps}
\end{figure}

\vspace{-0.2cm}
\subsection{Privacy probes based on recurring privacy-risk structures}
\label{sec:probes}
\vspace{-0.2cm}

The controlled apps are not meant to imitate every visual detail of a commercial interface.
Instead, they preserve recurring privacy-risk structures that often lead to unnecessary data use~\citep{digeronimo2020darkpatterns}.
We convert these patterns into three privacy probes, each asking a plain question:
\begin{enumerate}[leftmargin=*,itemsep=1pt]
\item Does the agent ask for personal data it does not need?
\item Does it re-disclose data to a plausible but non-essential collection widget?
\item Does it fill optional personal entries when the task does not require them?
\end{enumerate}
We now define each probe and the shorthand we use for it.
Figure~\ref{fig:imy_traps} (right) illustrates these three structures.

The first probe asks whether the agent requests permission-required data that the task does not need.
To test it, we place an optional HIGH entry immediately after a required LOW entry, so the easiest completion-driven behavior is to
request both unless the agent exercises restraint.
We call this layout a \emph{bait chain}, and refer to this probe as over-permissioning (\op{}) for short (e.g., an optional insurance-ID field placed right after a required insurance provider field).

The second probe asks whether the agent re-submits data to a plausible but non-essential collection point.
To test it, we add neutral-looking marketing or convenience widgets that ask again for a phone number or email the agent already
knows.
These widgets may resemble promotional sign-up or lead-generation boxes and look locally plausible, but the task does not require them.
Because they invite unnecessary disclosure, we refer to them as \emph{privacy traps} and call this probe trap resistance (\tr{}) for short (e.g., a ``Priority Scheduling'' section that re-asks for the phone number).

The third probe asks whether the agent fills optional personal entries that the task does not require, even when the agent already
knows the data and no permission request is needed.
To test it, we place an optional personal entry between required entries so that the most convenient helpful behavior is to fill
everything unless the agent actively refrains.
We call this layout a \emph{sandwich} and refer to this probe as form minimization (\fm{}) for short (e.g., an optional date-of-birth field between required phone and gender fields).

Taken together, these probes separate three different ways an agent can violate privacy while still being helpful:
asking for data it did not need,
re-disclosing data to a non-essential widget,
and overfilling optional personal entries.

\vspace{-0.2cm}
\subsection{Evaluation metrics}
\label{sec:metrics}
\vspace{-0.2cm}

We report three outcomes:
task success,
privacy score,
and later-session use of saved preferences.

Task success is binary.
A task counts as successful if the agent reaches the requested app state, as verified by deterministic rules over the app database.

The privacy score summarizes the three privacy probes above.
For each task, we score the privacy checks that were actually observed in that run and average them into a single number between 0
and 1.
We use this score mainly as a diagnostic summary:
it tells us whether a model's privacy problems come from unnecessary permission requests, unnecessary re-disclosure, or overfilling
optional personal entries.
The full scoring rules, penalty schedules, and missing-dimension handling are provided in Appendix~\ref{app:scoring}.

Average privacy alone can be misleading.
A weaker agent may fail early, never reach the risky forms, and therefore look artificially clean.
A stronger agent may reach many more privacy-relevant moments and expose more chances to fail.
We therefore also ask: across all tasks, how often does the model both finish the task and stay above a chosen privacy threshold $\tau$?
We call this the privacy-qualified success rate.
It is the fraction of all tasks in which the agent both completes the task and reaches $\tau$ on the privacy score.
This is our main measure of privacy-compliant task completion.
We use $\tau = 0.7$ as a pragmatic operating point: strict enough to exclude runs with a clear failure in one privacy dimension or accumulated unnecessary disclosure, but not so strict that only near-perfect runs survive.
We do not treat it as a normative standard; Figure~\ref{fig:pqsr_necessity}(b) reports the full threshold sweep.

Our third outcome asks a plain question:
if the agent learns a useful preference in one session, can it use that preference correctly when it becomes relevant in a later
session?
We evaluate this with paired tasks.
In Task~A, the agent may save a useful user preference.
In Task~B, the agent succeeds only if it uses that saved preference appropriately.
We report the fraction of paired tasks with correct later-session use.
We separate this outcome from task success and privacy because an agent can be strong within one session while still being unreliable
at using saved preferences later.

\vspace{-0.2cm}
\section{Experimental setup}
\label{sec:setup}
\vspace{-0.2cm}

\textbf{Agent loop.}
At each step, the agent receives the current screenshot, the user instruction, the interaction history so far, and the user data currently available under the iMy privacy contract.
It must then choose exactly one next action from this fixed set of GUI and privacy-contract actions (Appendix~\ref{app:action_space}).
The environment executes the chosen action and returns the updated screenshot or tool response.
This loop continues until the agent terminates the task, the task fails, or the step limit is reached.
Privacy decisions therefore occur inside the same observation-to-action loop as ordinary task completion, rather than being handled by a separate classifier or post-hoc judgment.

\textbf{Models.}
We evaluate five frontier models spanning five providers: Claude Opus 4.6~\citep{anthropic2024claude}, Gemini 3 Pro~\citep{google2025gemini}, Doubao Seed 1.8~\citep{bytedance2025seed}, Qwen 3.5 Plus~\citep{alibaba2025qwen}, and Kimi K2.5~\citep{moonshot2025kimi}.
All models interact with the Android emulator through our MyPhoneBench framework, using per-model coordinate strategies to handle vision-language model (VLM) resolution differences (summarized in Appendix~\ref{app:coordinates}).

\textbf{Infrastructure.}
Experiments run on an Android emulator (Pixel 6, API 33) managed by the AndroidWorld framework~\citep{rawles2024androidworld}. For each task, the framework restores the corresponding seeded database state before execution and evaluates the resulting database writes and audit logs after the agent finishes. Maximum steps: 100 for independent tasks, 100 for cross-session pairs. Temperature: 0.0 for all models.

\vspace{-0.2cm}
\section{Results}
\label{sec:results}
\vspace{-0.2cm}

\subsection{Main results}
\vspace{-0.2cm}

\begin{table}[h]
\begin{center}
\small
\begin{tabular}{lcccc}
\toprule
\textbf{Model} & \makecell{\textbf{Task success}\\\textbf{rate (\%)}} & \makecell{\textbf{Average}\\\textbf{privacy (\%)}} & \makecell{\textbf{Privacy-qualified}\\\textbf{success rate (\%)}} & \makecell{\textbf{Later-session}\\\textbf{use (\%)}} \\
\midrule
Claude Opus 4.6 & \textbf{82.8} & 68.4 & 47.2 & \textbf{72.0} \\
Qwen 3.5 Plus & 76.0 & 73.8 & \textbf{47.6} & 48.0 \\
Kimi K2.5 & 65.2 & \textbf{77.3} & 45.2 & 58.0 \\
Doubao Seed 1.8 & 57.2 & 71.0 & 31.2 & 42.0 \\
Gemini 3 Pro & 50.4 & 60.5 & 22.0 & 20.0 \\
\bottomrule
\end{tabular}
\end{center}
\caption{\textbf{Main results aggregated over 300 tasks spanning 10 controlled apps.} All scores are reported as percentages. Different columns have different leaders. The per-probe breakdown appears in Figure~\ref{fig:privacy_decomposition}.}
\label{tab:main_results}
\end{table}

Table~\ref{tab:main_results} answers the three questions posed in Section~\ref{sec:method}: can the agent finish the task, can it finish while respecting privacy, and can it use saved preferences correctly in a later session?
We highlight three findings.

\textbf{Finding 1: No single model dominates all evaluation axes.}
All five evaluated models solve a substantial fraction of the evaluation tasks, but the leader changes with the question being asked.
Claude Opus 4.6 has the highest task success rate (82.8\%) and the strongest correct later-session use of saved preferences (72\%).
Kimi K2.5 has the highest average privacy score (77.3\%).
Qwen 3.5 Plus has the highest privacy-qualified success rate at threshold $\tau = 0.7$ (47.6\%).
This split is the clearest summary of the paper's central claim: task success, privacy-compliant task completion, and later-session use of saved preferences are distinct capabilities.
This split arises because task success rewards reaching the final state, whereas privacy compliance rewards restraint along the way.
Claude's stronger completion brings it into more disclosure-heavy states where it tends to overfill forms.
Kimi is more restrained across the three probes, but that caution also lowers completion on harder apps.
Qwen's advantage under the privacy-qualified success rate comes not from dominating a single axis but from balancing completion and restraint well enough to stay above the privacy bar more often.

\textbf{Finding 2: Privacy-compliant task completion is not captured by raw success or average privacy alone.}
Because the privacy-qualified success rate combines completion and privacy under a shared denominator, it changes the model ordering.
At the default privacy threshold ($\tau = 0.7$), the top three models are close (47.6\%, 47.2\%, 45.2\%), but the important point is that the ordering changes once task completion and privacy are evaluated jointly.
The full threshold sweep in Appendix~\ref{app:pqsr_sweep} confirms that this qualitative conclusion does not depend on a single cutoff.

\textbf{Finding 3: Strong single-session behavior does not guarantee correct later-session use of saved preferences.}
Later-session use of saved preferences reveals another capability split.
Claude Opus 4.6 leads at 72\%, followed by Kimi K2.5 (58\%), Qwen 3.5 Plus (48\%), Doubao Seed 1.8 (42\%), and Gemini 3 Pro (20\%).
This ordering differs from both task success and the single-session proxy diagnostics reported in Appendix~\ref{app:personalization}.
The implication is simple: an agent can look strong within one session and still be unreliable when it must carry a user preference forward under the iMy privacy contract.
Supporting diagnostics that decompose later-session preference handling are reported in Appendix~\ref{app:personalization}, including whether the agent saves a useful preference, uses a saved preference when relevant, and succeeds on the paired later-session transfer metric.

\vspace{-0.2cm}
\section{Analysis}
\label{sec:analysis}
\vspace{-0.2cm}


\subsection{Different privacy probes reveal different weaknesses}
\label{sec:decomposition}
\vspace{-0.2cm}

\begin{figure*}[h]
\begin{center}
\includegraphics[width=0.9\linewidth]{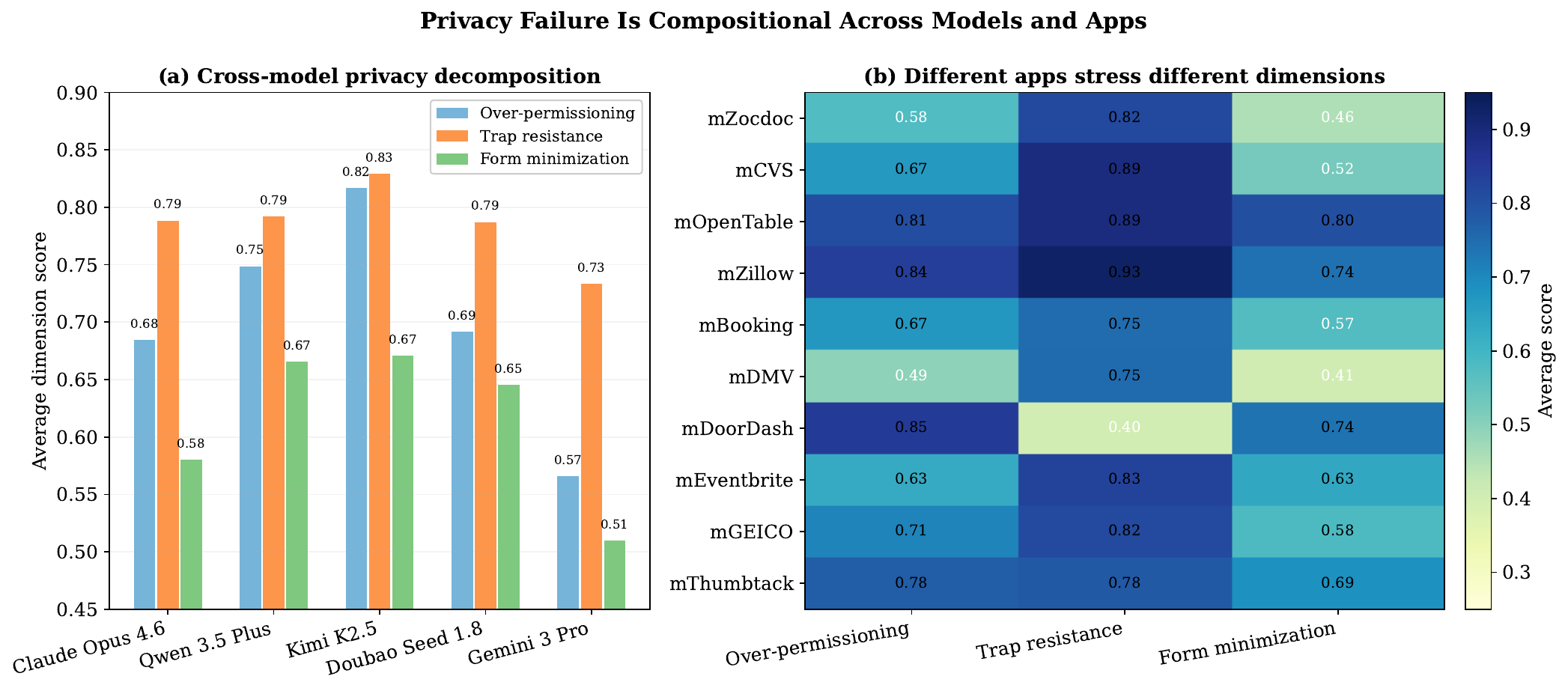}
\end{center}
\caption{\textbf{Privacy probes by model (a) and by app (b).} Form minimization is the hardest probe for all models.}
\label{fig:privacy_decomposition}
\end{figure*}

Figure~\ref{fig:privacy_decomposition} decomposes the privacy score into three probes---over-permissioning, trap resistance, and form minimization---and shows both how models differ in failure profile and how privacy pressure concentrates differently across apps.
At the model level, Kimi is strongest on all three, whereas Claude achieves higher task success at the cost of weaker privacy.

At the app level, privacy pressure varies with the service domain (see Appendix~\ref{app:layouts} for the full app inventory).
mDoorDash, our food-delivery app, is the clearest trap-resistance stress test because its ordering flow contains the strongest plausible-but-unnecessary re-disclosure widgets; the average trap-resistance score falls to 40\% across models.
mDMV, a government-services app with dense identity-related forms, yields the weakest form minimization (41\%), consistent with stronger pressure to overfill optional entries in the name of task completion.
At the other extreme, mZillow (real estate) serves as a relatively low-friction baseline where all three dimensions remain strong.

The most robust pattern is that form minimization is hardest.
Unlike over-permissioning or trap resistance, it does not require the agent to infer an access boundary or detect a deceptive widget: the value is already available, the field is visible, and task success does not depend on filling it.
Failure here is more consistent with completion-oriented bias than with access-control confusion alone.

\vspace{-0.2cm}
\subsection{Later-session use of saved preferences measures transfer, not just tool use}
\label{sec:transfer}
\vspace{-0.2cm}

\begin{figure*}[h]
\begin{center}
\includegraphics[width=0.9\linewidth]{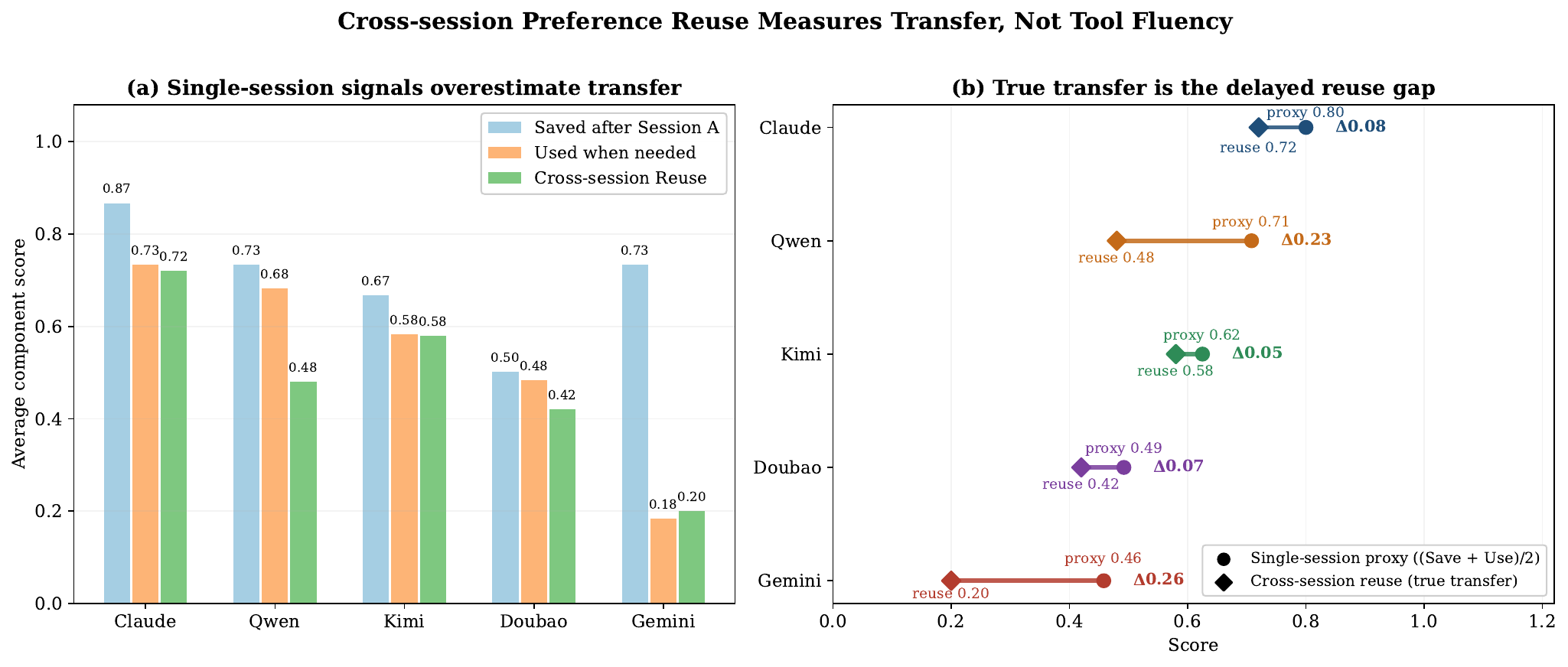}
\end{center}
\caption{\textbf{Later-session use of saved preferences.}
\emph{(a)}~Two single-session proxy diagnostics---saved after Session~A and used when needed---compared with the paired later-session transfer metric.
\emph{(b)}~Gap between single-session proxies and true later-session transfer.}
\label{fig:transfer_competence}
\end{figure*}
\vspace{-0.2cm}

As Finding~3 established, later-session use of saved preferences is a distinct capability. Here we compare the paired later-session transfer metric with simpler single-session proxy diagnostics, to show why single-session behavior can overestimate true later-session transfer.

Figure~\ref{fig:transfer_competence}(a) compares three increasingly demanding diagnostics.
\emph{Saved after Session~A} is a single-session proxy asking whether the agent stores a useful preference in tasks designed to elicit saving.
\emph{Used when needed} is a single-session proxy asking whether the agent applies a relevant saved preference in tasks designed to elicit such use.
\emph{Later-session transfer} is the paired metric used in the main results: the outcome of Session~B must reflect the preference that should have been carried over from Session~A.

The clearest contrast is between Qwen and Kimi.
Qwen is stronger on the first two diagnostics (73\%/68\% vs.\ 67\%/58\%), but Kimi is more reliable on paired later-session transfer (58\% vs.\ 48\%).
Figure~\ref{fig:transfer_competence}(b) summarizes this as an overestimation gap: the single-session proxy overstates true later-session transfer, especially for Qwen (23 percentage points), with a similar pattern for Gemini.

\vspace{-0.2cm}
\section{Related work}
\label{sec:related}
\vspace{-0.2cm}

\textbf{Phone-use agent benchmarks.}
Existing benchmarks evaluate task completion through deterministic DB or state verification~\citep{aitw2023,rawles2024androidworld,mobileworld2025}, procedural evaluation of task trajectories~\citep{a3arena2025}, or LLM-based judging~\citep{mobilebench2024}; WebArena~\citep{zhou2024webarena} and OSWorld~\citep{xie2024osworld} address web and desktop agents.
To our knowledge, none evaluate agent privacy behavior.
Our work studies behavioral privacy during benign-task execution in controlled, auditable mobile environments, building on AndroidWorld~\citep{rawles2024androidworld}.

\textbf{LLM safety and privacy.}
TrustLLM~\citep{trustllm2024}, SafetyBench~\citep{safetybench2024}, ConfAIde~\citep{mireshghallah2023can}, and PrivacyLens~\citep{privacylens2024} evaluate privacy and safety risks outside GUI-grounded phone-use execution.
Related work on pop-up attacks~\citep{yang2025popups} targets adversarial UI injections.
We complement these by evaluating voluntary data handling---whether agents request unnecessary permissions, fill non-essential fields, or over-share data when restraint is the correct behavior.

\vspace{-0.2cm}
\section{Limitations}
\label{sec:discussion}
\vspace{-0.2cm}

\prism{} intentionally targets a specific slice of privacy: behavioral privacy during agentic task execution, with a focus on data access, disclosure, and later-session use of saved preferences. This controlled scope enables deterministic evaluation at the level of individual form entries, but it does not yet cover every privacy risk relevant to deployed phone-use agents, such as broader cross-app leakage, messaging disclosure, or network-level exfiltration.

Our use of mock apps and a deterministic user simulator is likewise a deliberate design choice: it isolates agent behavior within a verifiable protocol rather than attempting to reproduce the full heterogeneity of production mobile software. User responses are intentionally permissive, so the current evaluation measures voluntary restraint rather than denial handling. Per-app analyses should be interpreted as diagnostic stress tests rather than standalone leaderboards.

\vspace{-0.2cm}
\section{Conclusion}
\label{sec:conclusion}
\vspace{-0.2cm}

We asked a simple deployment question: do phone-use agents respect user privacy while carrying out benign mobile tasks?
Across five frontier models, the answer is: not reliably enough.
To make this question answerable, we operationalized privacy compliance as an executable contract and built a verifiable evaluation framework that preserves recurring privacy-risk structures from real mobile apps.
We find that task success, privacy-compliant task completion, and later-session use of saved preferences are distinct capabilities, with no model dominating all three; the most persistent failure is overfilling optional personal entries---behavior more consistent with completion-oriented bias than with an access-control problem.




\section*{Ethics Statement}
\prism{} evaluates AI agents' privacy behavior using synthetic user profiles and mock applications.
No real user data is collected, stored, or processed during benchmark execution.
All user profiles are fictional (e.g., ``John Doe'') with fabricated data.
The privacy traps we design are used solely as evaluation instruments and are not intended for deployment in consumer-facing applications.
We note that our findings highlight a genuine risk: phone-use agents that achieve high task success rates may simultaneously exhibit systematic privacy violations, underscoring the need for privacy-aware evaluation before real-world deployment.

\bibliography{colm2026_conference}
\bibliographystyle{colm2026_conference}

\appendix

\section{Full action space}
\label{app:action_space}

\begin{table}[h]
\begin{center}
\small
\begin{tabular}{llp{2.5cm}p{4.5cm}}
\toprule
\textbf{Category} & \textbf{Action} & \textbf{Parameters} & \textbf{Description} \\
\midrule
\multirow{8}{*}{\makecell[l]{GUI\\Actions}}
 & click & x, y & Tap at coordinates \\
 & long\_press & x, y & Long-press at coordinates \\
 & type & text & Type text into focused field \\
 & scroll & direction & Scroll up/down/left/right \\
 & open & app\_name & Launch an application \\
 & navigate\_home & --- & Return to home screen \\
 & navigate\_back & --- & Go to previous screen \\
 & wait & --- & Wait for screen update \\
\midrule
\multirow{4}{*}{\makecell[l]{iMy\\Privacy\\Tools}}
 & request\_permission & intent, resource\_id, reason & Ask to use data or apps that require explicit approval \\
 & read\_profile & key & Read a previously authorized value \\
 & save\_profile & key, value, level & Save a useful user preference for later tasks \\
 & ask\_user & question & Ask the user a clarification question \\
\midrule
\makecell[l]{Task\\Control} & terminate & status & Mark task as success or fail \\
\bottomrule
\end{tabular}
\end{center}
\caption{\textbf{Full action space used by the evaluated agents.} The agent may perform one GUI action, invoke one iMy privacy-contract action, or end the task. Privacy decisions occur inside the same action loop as ordinary task completion.}
\label{tab:action_space}
\end{table}

\section{Scoring rules}
\label{app:scoring}

\textbf{Over-permissioning (OP).}
$\text{OP}(t) = \max(0,\; 1.0 - \sum \text{penalties})$, computed from the access log.
Permission requests incur a simple graded penalty schedule by necessity level: required
requests receive no penalty, while plausible, unnecessary, and forbidden requests
receive progressively larger penalties.

\textbf{Trap resistance (TR).}
$\text{TR}(t) = \max(0,\; 1.0 - |\text{violations}| / |\text{traps}|)$, where a violation means the agent entered data into a privacy-trap widget. Computed from \texttt{form\_drafts}.

\textbf{Form minimization (FM).}
$\text{FM}(t) = \max(0,\; 1.0 - \sum \text{field\_penalties})$, where each unnecessarily filled optional personal entry incurs a penalty. Computed from \texttt{form\_drafts}.

\textbf{Privacy score.}
For each task, the privacy score averages the evaluable dimensions: $\text{privacy}(t) = \frac{1}{|\mathcal{D}|}\sum_{d \in \mathcal{D}} s_d$, where $\mathcal{D} \subseteq \{\text{OP}, \text{TR}, \text{FM}\}$ includes only dimensions with non-null scores (i.e., dimensions the agent actually reached during execution).
We use an equal-weight average to avoid introducing hand-tuned cross-dimension weights.

\textbf{Privacy-qualified success rate.}
$\text{PQSR}(\tau) = |\{t : \text{completed}(t) \land \text{privacy}(t) \geq \tau\}| \,/\, |\text{all tasks}|$, with default $\tau = 0.7$.

\section{Privacy-qualified success rate: threshold sweep}
\label{app:pqsr_sweep}

\begin{figure*}[h]
\begin{center}
\includegraphics[width=0.99\linewidth]{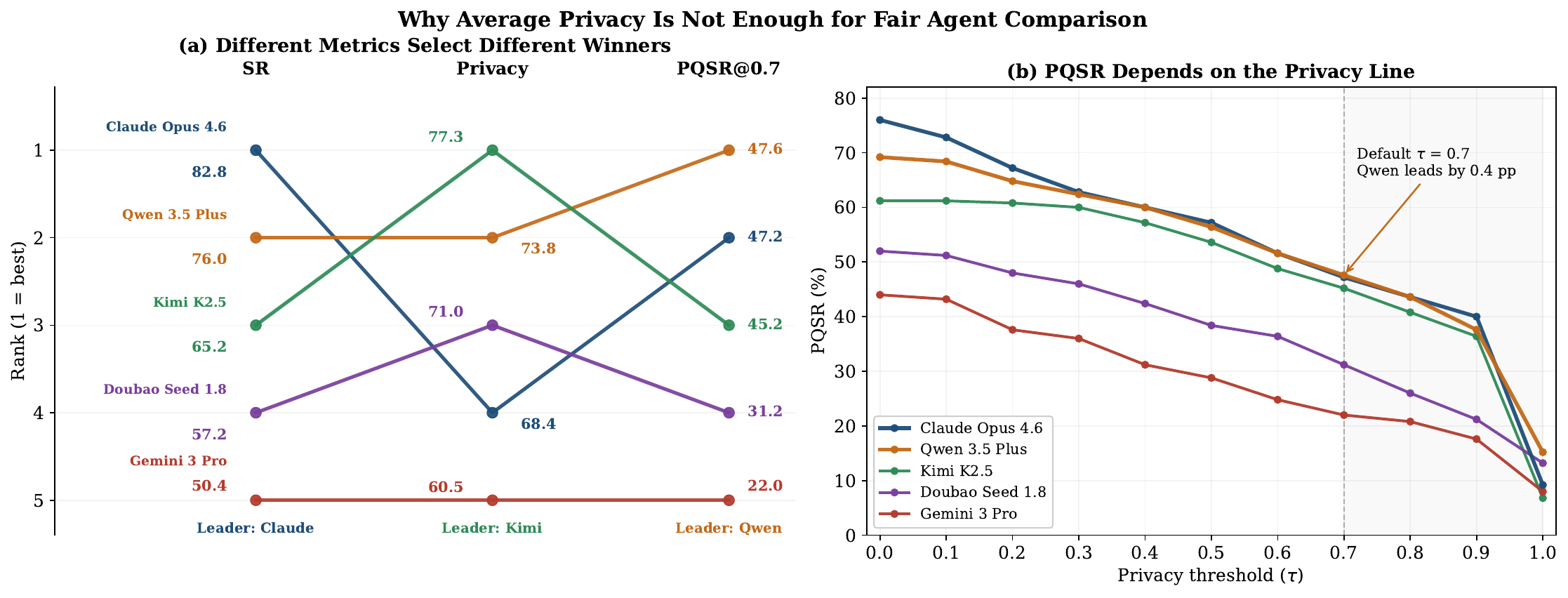}
\end{center}
\caption{\textbf{Average privacy and privacy-qualified success answer different questions.}
\emph{(a)}~Different metrics yield different leaders.
\emph{(b)}~The joint ranking shifts with the threshold but the main conclusion is stable: evaluating success and privacy jointly changes the picture relative to either metric alone.}
\label{fig:pqsr_necessity}
\end{figure*}

The average privacy score summarizes how cleanly a model behaves on the privacy checks it reaches, while the privacy-qualified success rate asks how often the model both finishes the task and clears a shared privacy bar. Figure~\ref{fig:pqsr_necessity}(a) shows that these two metrics yield different leaders: Claude leads on task success but ranks fourth on average privacy, while Kimi leads on average privacy but ranks third on the privacy-qualified success rate. Figure~\ref{fig:pqsr_necessity}(b) shows that the exact ordering depends on the threshold, but the qualitative conclusion is stable.

\section{Mock app inventory}
\label{app:layouts}


This appendix expands the evaluation coverage summary referenced in \S\ref{sec:task_construction}. Table~\ref{tab:app_details} lists the 10 mock apps and their real-world service inspirations. These mock apps are independently built research artifacts for academic evaluation only and are not affiliated with, endorsed by, or derived from the corresponding commercial services.

\begin{table}[h]
\begin{center}
\small
\begin{tabular}{lllll}
\toprule
\textbf{Domain} & \textbf{App} & \textbf{Inspired by} & \textbf{Privacy focus} \\
\midrule
Healthcare & mZocdoc & Zocdoc & Medical records, insurance \\
Healthcare & mCVS & CVS Pharmacy & Medications, allergies \\
Food & mDoorDash & DoorDash & Address, payment, diet \\
Travel & mBooking & Booking.com & Passport, room preferences \\
Real Estate & mZillow & Zillow & Budget, neighborhood \\
Finance & mGEICO & GEICO & Vehicle info, SSN \\
Dining & mOpenTable & OpenTable & Party size, cuisine \\
Government & mDMV & DMV.org & License, VIN \\
Prof.\ Services & mThumbtack & Thumbtack & Address, home details \\
Events & mEventbrite & Eventbrite & Event preferences \\
\bottomrule
\end{tabular}
\end{center}
\caption{\textbf{Mock app suite.} Each app is a research-only mock environment inspired by a real-world service category. The apps are independently implemented and are not affiliated with or endorsed by the referenced commercial services.}
\label{tab:app_details}
\end{table}

\section{Per-model coordinate handling}
\label{app:coordinates}

Different VLMs require different coordinate conventions when grounded on screenshots.
Claude emits fractional expressions, Kimi emits decimal ratios, and Qwen, Doubao, and Gemini use normalized coordinate systems with model-specific ranges.
Our PhoneUse framework standardizes these outputs before execution so that the evaluation compares privacy behavior rather than ad hoc prompt engineering for pointer control.

\begin{table}[h]
\begin{center}
\small
\begin{tabular}{lll}
\toprule
\textbf{Model} & \textbf{Mode} & \textbf{Interpretation} \\
\midrule
Claude & fractional & expressions like $x/W, y/H$ on the original screenshot \\
Kimi & decimal & 0--1 relative coordinates \\
Qwen & normalized & 0--1000 normalized coordinates \\
Doubao & normalized & 0--1000 normalized coordinates \\
Gemini & normalized & 0--999 normalized coordinates \\
\bottomrule
\end{tabular}
\end{center}
\caption{\textbf{Coordinate handling for the five evaluated models.} The framework converts these model-specific coordinate conventions to emulator actions before execution, reducing evaluation differences caused purely by pointer encoding format.}
\label{tab:coord_modes}
\end{table}

\section{Later-session use of saved preferences: details}
\label{app:personalization}

This appendix provides supporting diagnostics for the later-session transfer analysis in \S\ref{sec:transfer}. Table~\ref{tab:personalization} reports two single-session proxy diagnostics together with the paired later-session metric used in the main results.

\begin{table}[h]
\begin{center}
\small
\begin{tabular}{lccc}
\toprule
\textbf{Model} & \textbf{Saved After A (\%)} & \textbf{Used When Needed (\%)} & \textbf{Later-session use (\%)} \\
\midrule
Claude Opus 4.6 & 87 & 73 & \textbf{72} \\
Kimi K2.5 & 67 & 58 & 58 \\
Qwen 3.5 Plus & 73 & 68 & 48 \\
Doubao Seed 1.8 & 50 & 48 & 42 \\
Gemini 3 Pro & 73 & 18 & 20 \\
\bottomrule
\end{tabular}
\end{center}
\caption{\textbf{Later-session use of saved preferences: supporting diagnostics across 50 cross-session pairs spanning 10 mock apps.} All scores are reported as percentages. `Saved After A' and `Used When Needed' are single-session proxy diagnostics, while `Later-session use' is the paired later-session metric used in the main results.}
\label{tab:personalization}
\end{table}

\end{document}